\documentclass[twocolumn,preprintnumbers,amsmath,amssymb]{revtex4}

\usepackage{graphicx}
\usepackage{dcolumn}
\usepackage{bm}
\newcommand\beq{\begin{eqnarray}}
\newcommand\eeq{\end{eqnarray}}

\newcommand\Fig[1]{Fig.~\ref{fig:#1}}
\newcommand\bal{\begin{align}}
\newcommand\eal{\end{align} }
\newcommand\Eq[1]{Eq.~\ref{eq:#1}}

\newcommand{\mybar}[1]%
        {\kern 0.6pt\overline{\kern -0.6pt#1\kern -0.6pt}\kern 0.6pt}

\begin{document}

\preprint{}

\title{$N$-body Efimov states from two-particle noise}

\author{Amy N. Nicholson}%
 \email{amynn@umd.edu}
\affiliation{Maryland Center for Fundamental Physics, Department of Physics, University of Maryland, College Park MD 20742-4111
}%


\begin{abstract}
The ground state energies of universal $N$-body clusters tied to Efimov trimers, for $N$ even, are shown to be encapsulated in the statistical distribution of two particles interacting with a background auxiliary field at large Euclidean time when the interaction is tuned to the unitary point. Numerical evidence that this distribution is log-normal is presented, allowing one to predict the ground state energies of the $N$-body system.
\end{abstract}

\maketitle

While noisy correlators are generally regarded as an impediment for Monte Carlo calculations, there is often a physical mechanism underlying the appearance of noise. Understanding the form of the statistical distribution can lead not only to better methods for extracting reliable quantities from numerical calculations, but also to insight into the physics itself. In this work, we illustrate this principle using a correlator for two particles with an interaction tuned to give a bound state at threshold, called the unitary point. From the distribution of this correlator we are able to extract the energies of $2N$-body Efimov states, deeply bound universal systems of bosons or distinguishable fermions tuned to unitarity, which are tied to Efimov trimers \citep{V1970563,Efimov:1971zz}. Efimov physics has enjoyed a resurgence of interest among the atomic, nuclear, and condensed matter communities due to progress in the theoretical understanding of Efimov physics \citep{Bedaque:1998kg,Bedaque:1998km,Mohr:2003du,Platter:2004qn} as well as advances in ultracold atom experiments, particularly with recent experimental evidence for three- and four-body Efimov states displaying universal characteristics \citep{2006Natur.440..315K,PhysRevLett.101.203202,2009NatPh...5..586Z,PhysRevLett.102.165302,PhysRevLett.102.140401,Pollack18122009}, meaning their low-energy behavior is independent of the details of the interaction. However, theoretical information about the existence and properties of higher-body systems has been limited to that from direct measurements of $N$-body states using non-perturbative numerical methods \citep{2006PhRvA..74f3604H,vonStecher:2009qw,2011PhRvL.107t0402V,2011PhRvA..84e2503G}.

Recent lattice studies of many-fermion systems at unitarity have shown that the correlators display distributions with log-normal characteristics \citep{Endres:2011jm,Endres:2011mm}. 
Using our two-body correlator we establish a deep connection between Efimov physics and the log-normal distribution. We then present lattice data which indicates that the correlator is log-normal to within $2\%$, and use this knowledge to derive an analytical prediction for the energies of $2N$-body Efimov states. Finally, we compare our results to those from numerical calculations.

To begin, we will define the two-body correlation function of interest and study its distribution by calculating the moments. The Lagrangian consists of two degenerate flavors of nonrelativistic particles \footnote{Throughout the paper we will consider systems of degenerate, distinguishable particles; however, as we are interested only in the energy levels of such systems, the results are equally applicable to systems of one-component bosons.} interacting through a point interaction with coupling $\kappa$,
\beq
\mathcal{L} = \psi^{\dagger}(\partial_{\tau}-\nabla^2/2M) \psi + \kappa^2 (\psi^{\dagger}\psi)^2 \, .
\eeq
Performing a Hubbard-Stratonovich transformation gives the Euclidean path integral in terms of an auxiliary field, $\phi$,
\beq
Z&=&\int \mathcal{D}\psi^{\dagger}\mathcal{D}\psi \mathcal{D}\phi e^{-\int d\tau d^3x \left( \psi^{\dagger} K \psi + \frac{1}{2} \phi^2 \right)} \, , \cr
K &=& (\partial_{\tau}-\nabla^2/2M) + \kappa \phi \, .
\eeq
The two-particle correlation function is given by
\beq
\label{eq:C2}
C_2(T)  = \langle \Psi(T) \Psi^{\dagger}(0) \rangle \, ,
\eeq
where $\Psi(\tau) = \int dx_1 dx_2 A(x_1,x_2) \psi_{\uparrow}(x_1,\tau) \psi_{\downarrow}(x_2,\tau)$ annihilates a two-body state, with wavefunction $A$ at time $\tau$, which has non-zero overlap with the two-body ground state. 
Integrating out the $\psi$ fields gives the correlation function as a path integral over $\phi$ field configurations only,
\beq
\label{eq:part_func}
C_2(T) &=& \langle S_2(\phi,T) \rangle_{\phi} \cr 
&=& \frac{1}{Z_{\phi}} \int \mathcal{D}\phi (\det K)^2 \mathcal{S}_2(\phi,T) e^{-\int d\tau d^3x  \frac{1}{2} \phi^2 } \, , \cr
Z_{\phi} &\equiv&  \int \mathcal{D}\phi (\det K)^2 e^{-\int d\tau d^3x  \frac{1}{2} \phi^2 } \, ,
\eeq
where $\mathcal{S}_2(\phi,T)$ is the two-particle propagator from Euclidean time $\tau = 0$ to $T$ on a given background field, $\phi$, and we have one power of $\det K$ for each flavor.
Using open temporal boundary conditions, one may show that $\det K$ is a constant independent of $\phi$, and therefore may be disregarded \citep{Endres:2011er}. The use of open boundary conditions is justified so long as we restrict our arguments to zero temperature (large Euclidean time). 

By inserting a complete set of energy eigenstates, $| n \rangle $, in \Eq{C2}, one may show that for large Euclidean time, $T$, $C_2$ will be dominated by the ground state,
\beq
C_2(T) = \sum_n \langle \Psi(0) | n \rangle e^{-E_n^{(2)} T} \langle n | \Psi(0) \rangle \underset{T \to \infty}{\longrightarrow} \mathcal{Z}_2 e^{-E_0^{(2)} T}  ,
\eeq
where $E_0^{(2)}$ is the ground state energy of the two-body system and $\mathcal{Z}_2$ gives the overlap of the operator $\Psi$ with the ground state. 

We determine the second moment of the correlator by investigating the expectation value of the square of the operator,
\beq
\mathcal{M}_2(T) &\equiv& \langle | S_2(\phi,T)|^2 \rangle_{\phi} \cr
&=& \frac{1}{Z_{\phi}} \int \mathcal{D}\phi ( \mathcal{S}_2[\phi,T])^2 e^{-\int d\tau d^3x  \frac{1}{2} \phi^2 } \, ,
\eeq
where we have dropped the determinant for the reason discussed above. Following the analysis of Lepage \citep{Lepage:1989hd}, we may interpret this quantity physically by noting that it is the correlator for the product of two 2-body propagators, corresponding to a four-particle system. Because there is no (anti-)symmetrization between the four (fermions)bosons, each particle must correspond to a different flavor. Note that because the partition function does not depend on the fermion determinant, it is unchanged when the number of flavors is increased, so this correlation function properly describes a physical four-body system. 

For the large Euclidean time limit we have $\mathcal{M}_2(T)  \underset{T \to \infty}{\longrightarrow} \mathcal{Z}_4 e^{- E_0^{(4)} T}$, where $E_0^{(4)}$ is the ground state energy of the four-particle, four-flavor system, and $\mathcal{Z}_4$ gives the overlap of $\Psi^2$ with the four-particle ground state. We may generalize this argument for all moments, 
\beq
\label{eq:genmoments}
\mathcal{M}_{N}(T)  \underset{T \to \infty}{\longrightarrow} \mathcal{Z}_{2N} e^{- E_0^{(2N)} T} \, ,
\eeq
where $E_0^{(2N)}$ is the ground state energy of the $2N$-particle state in the $2N$-flavor theory and $\mathcal{Z}_{2N}$ represents the overlap of $\Psi^N$ with the $2N$-particle ground state, which is assumed to be non-zero but may be arbitrarily small. 

Let us now consider the spectrum of these systems in the unitary limit, where the two-body ground state energy is zero by construction. The three-body system exhibits the well-known Efimov effect \citep{V1970563,Efimov:1971zz}, consisting of a series of bound trimers whose energies are separated by a factor of $\sim 515$. The ground-state is stabilized against collapse by an effective 3-body interaction which becomes relevant for low energy physics. Hence, the addition of a third particle to the conformally invariant two-particle system introduces an energy scale, $\Lambda_E$, to which the 3-body ground-state binding energy may be related, $E_0^{(3)} = - a_3 \Lambda_E$.

Theoretical studies \citep{vonStecher:2009qw,2006PhRvA..74f3604H,2007EPJA...32..113H,2010PhRvA..81e2709S,2008arXiv0810.3876V,2012arXiv1202.0167D} as well as recent experiments \citep{PhysRevLett.102.140401,Pollack18122009} indicate that the spectrum of the four-body, four-flavor theory consists of a set of two bound tetramers tied to each Efimov trimer. 
Provided the UV physics is fixed such that the ground state is sufficiently far from the cutoff, the ratio of the lowest four-body state to the lowest Efimov trimer has been shown to be a universal constant \citep{Platter:2004qn,2006PhRvA..74f3604H,vonStecher:2009qw,2007EPJA...32..113H,2008arXiv0810.3876V,2012arXiv1202.0167D}. This implies that there is no relevant four-body scale for this system, so $\Lambda_E$ remains the only scale, giving $E_0^{(4)} = - a_4 \Lambda_E$.

One may postulate that no new relevant scales will emerge for the $N$-body system at unitarity, and numerical evidence for up to $N=40$ suggests that this is true \citep{2006PhRvA..74f3604H,vonStecher:2009qw,2011PhRvA..84e2503G}. Accordingly, the energy of the $N$-body system will be $E_{0}^{(N)} = - a_N \Lambda_E$. Provided the ground state energy of the system obeys this scaling property, there will be at least one $N$-body state tied to each Efimov trimer, thus, the excited states of the $N$-body system must obey the same discrete scale invariance as the three-body system. Consequently, the calculation of the $a_{N}$ determines the entire spectrum for this series of states. Additionally, there may be multiple states tied to each Efimov trimer, as seen in the four-body system. Here we only concentrate on the lowest of these possible states.

We emphasize that while we only consider the ground states of these systems, provided the energy cutoff is much larger than the energy per particle of the $N$-body ground state, these states will still exhibit universal characteristics, and nonuniversal effects will be small. Even given the discrete scale invariance of these systems, the cutoff may be up to $\sim 515$ times the ground state energy per particle, so it is possible in practice to considerably reduce nonuniversal corrections to the ground states, as numerical calculations of these systems have indicated \citep{2006PhRvA..74f3604H,vonStecher:2009qw}. 

Returning to the probability distribution, we may combine the above relation between energies at unitarity with the moments of the distribution, giving
\beq
\label{eq:unitarymoments}
\mathcal{M}_{N}^{\mbox{\small unit}}(T) \to \mathcal{Z}_{2N} e^{a_{2N} \Lambda_E T}
\eeq

Now we shall discuss the observed behavior of the distribution using lattice results. In \citep{Endres:2011jm} it was shown that lattice calculations of strongly interacting non-relativistic systems tend to display heavy-tailed distributions at large Euclidean time. In particular, it was found that the distributions are approximately log-normal (LN), i.e. that the logarithm of the quantity of interest obeys the normal distribution. It was also shown that an expansion around LN, called the cumulant expansion, can be used to extract a reliable mean for the correlator. Using this method, the moments of $\ln C$ are calculated and the correlator is given by $\ln C(T) = \lim_{N_{\kappa} \to \infty} \ln C^{(N_\kappa)}(T) $, where
\begin{eqnarray}
\label{eq:cumulantexp}
\ln C^{(N_\kappa)}(T) \equiv \sum_{n=1}^{N_{\kappa}} \frac{ \kappa_n(T) }{n!} \, .
\label{eq:cumulant_expansion}
\end{eqnarray}
Here, $\kappa_n(T)$ is the $n$-th cumulant of the distribution for $\ln C(\phi,T)$, and the expansion may be cut off for finite $N_{\kappa}$ provided the distribution is close enough to LN that the series converges. For the LN distribution, $\kappa_n = 0$ for $n \geq 3$, so the expansion may be cut off exactly at $N_\kappa = 2$. 

\begin{figure}
\includegraphics[width=0.9 \linewidth]{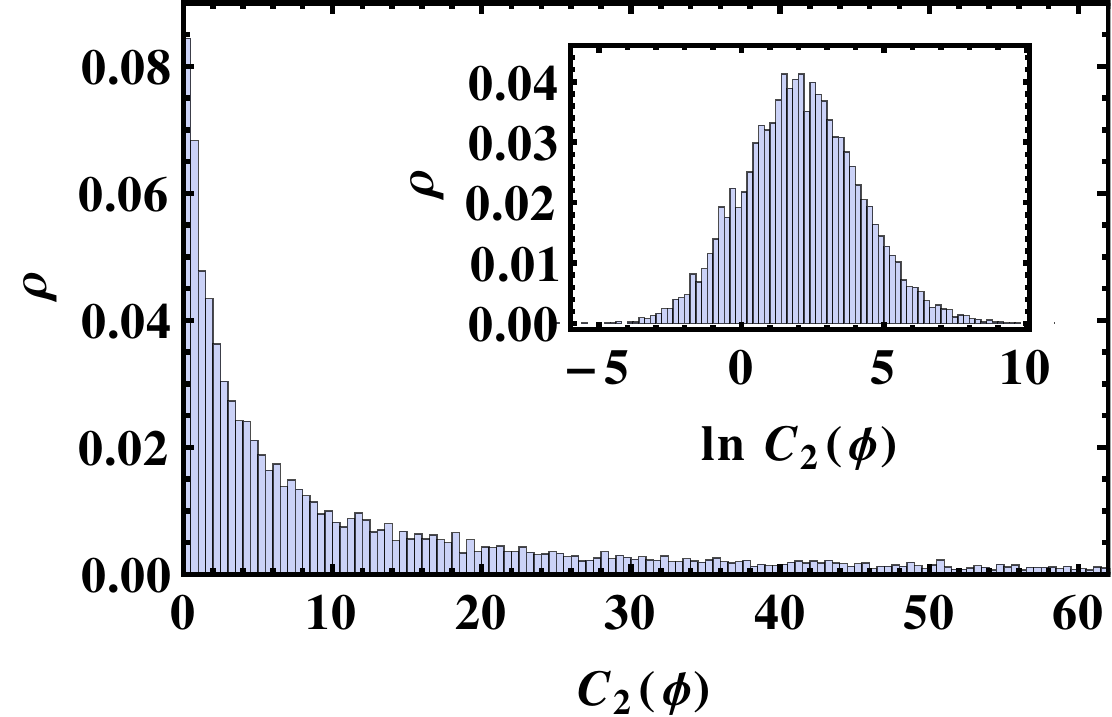}
\caption{\label{fig:dist}Histogram of the two-body correlator at large Euclidean time, $T$. Inset: Histogram of the logarithm of the correlator at the same value of $T$. }
\end{figure}

A histogram of the two-particle correlator at unitarity is shown in \Fig{dist} for large Euclidean time. The distribution of the logarithm of the correlator is also shown in the inset. The data was generated using the lattice method developed in \citep{Endres:2011er}. A set of momentum-dependent interactions are tuned to systematically correct for lattice artifacts, bringing us closer to the unitary point by effectively setting the range of the interaction and the first few shape parameters to zero. Preliminary calculations have shown that an $L=16$ box is sufficient to eliminate finite volume effects for the three-body state \citep{Nicholson:2012}, which is the largest of the $N$-body bound states and therefore most susceptible to finite volume errors. Finally, there is a hard momentum cutoff in our formulation. We find that in the absence of an explicit three-body interaction the resulting energy cutoff is approximately 335 times the three-particle ground state energy per particle. 

We see that this correlator exhibits a very heavy tail, while the logarithm of the correlator appears to be Gaussian, characteristic of the LN distribution. To understand physically why this occurs, let us compare the moments derived above to those of the LN distribution.
For the LN distribution the $n$-th moment is
\beq
\label{eq:lngenmoments}
\mathcal{M}_n^{\mbox{\tiny LN}} = e^{n \mu + \frac{1}{2} n^2 \sigma^2}
\eeq
where $\mu, \sigma$ are the mean and standard deviation, respectively, of the logarithm of the correlator. 

We may extract values for these parameters by setting \Eq{genmoments} equal to \Eq{lngenmoments} for $n=N=1$ and $n=N=2$. For the moment we will ignore the overlap factors $\mathcal{Z}_{2N}$ and consider only the $T$-dependence of the parameters. The addition of the overlap factors will be discussed at a later point. The result then becomes $\mu = \frac{1}{2} E_0^{(4)} T \, , \sigma^2 = - E_0^{(4)} T$. Thus, for the special case of unitarity, where $E_0^{(2)} = 0$, we find that $\mu$ and $\sigma$ are no longer independent. This implies that there is only one scale controlling all moments of the distribution. The moments become,
\beq
\label{eq:logmoments}
\mathcal{M}_n^{\mbox{\tiny LN}} \propto e^{-\frac{1}{2}n(n-1) E_0^{(4)}T} = e^{\frac{1}{2}n(n-1) a_4 \Lambda_ET}\, .
\eeq

A comparison of \Eq{unitarymoments} and \Eq{logmoments}, shows that the moments given by the LN distribution display the same scaling behavior with $\Lambda_E, T$ as the moments predicted by the Efimov spectrum. Note that the LN distribution is the maximum entropy distribution of its class: once the parameters $\mu$ and $\sigma$ have been fixed by the energies $E_0^{(2)}$ and $E_0^{(4)}$, the distribution should be LN in the absence of further constraints on the higher moments.

Finally, assuming the distribution of the two-body correlator is exactly LN, we can deduce the energies of all $2N$-body states by equating \Eq{genmoments} with \Eq{logmoments}:
\beq
\label{eq:LNenergies}
E_0^{(2N)} = \frac{1}{2}N(N-1) E_0^{(4)} 
\eeq
We see that the $N$-dependence of the energy relation is equivalent to the number of pairwise interactions between dimers. 

For an approximately LN distribution one could imagine systematically improving this relation by numerically calculating third and higher cumulants of $\ln C_2$, using this information to correct the distribution, for example via the principle of maximum entropy, and recalculating the moments of $C_2$. We would like to emphasize that this approach does not simply translate a difficult many-body problem into an equally difficult problem of extracting large moments of a distribution, but rather small moments of $\ln C_2$.

\begin{figure}
\includegraphics[width=0.9 \linewidth]{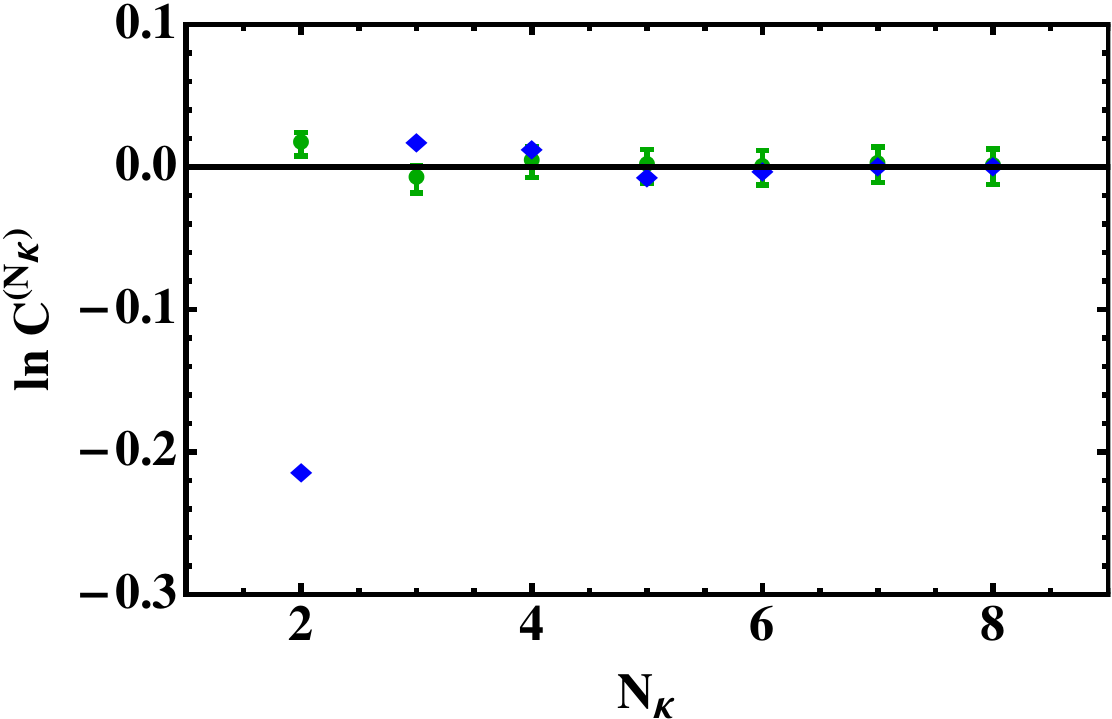}
\caption{\label{fig:exp}(Color online) Deviation between the cumulant expansion (\Eq{cumulantexp}) cut off at $N_{\kappa}$ and the converged value. The green circles represent the lattice data for the two-body correlator tuned to unitarity on a $16^3 \times 1000$ lattice, while the blue diamonds are from a mock distribution created to reproduce moments corresponding to the energies found in \citep{vonStecher:2009qw}. }
\end{figure}

Remarkably, to within a few percent, we find that the distribution is indeed LN. 
In \Fig{exp} (green circles) we plot the cumulant expansion for the lattice data. Recall that for the LN distribution, this expansion converges at $N_{\kappa}=2$. Thus, the discrepancy between the expansion cut off at $N_{\kappa}=2$ and the result after convergence may be used to quantify how close to LN a distribution is. We find that our lattice data is LN to within $\sim 2\%$. This small discrepancy is likely due to the finite time extent used and sensitivity to lattice artifacts for large moments of $C_2$.

\begin{figure}
\includegraphics[width=0.9 \linewidth]{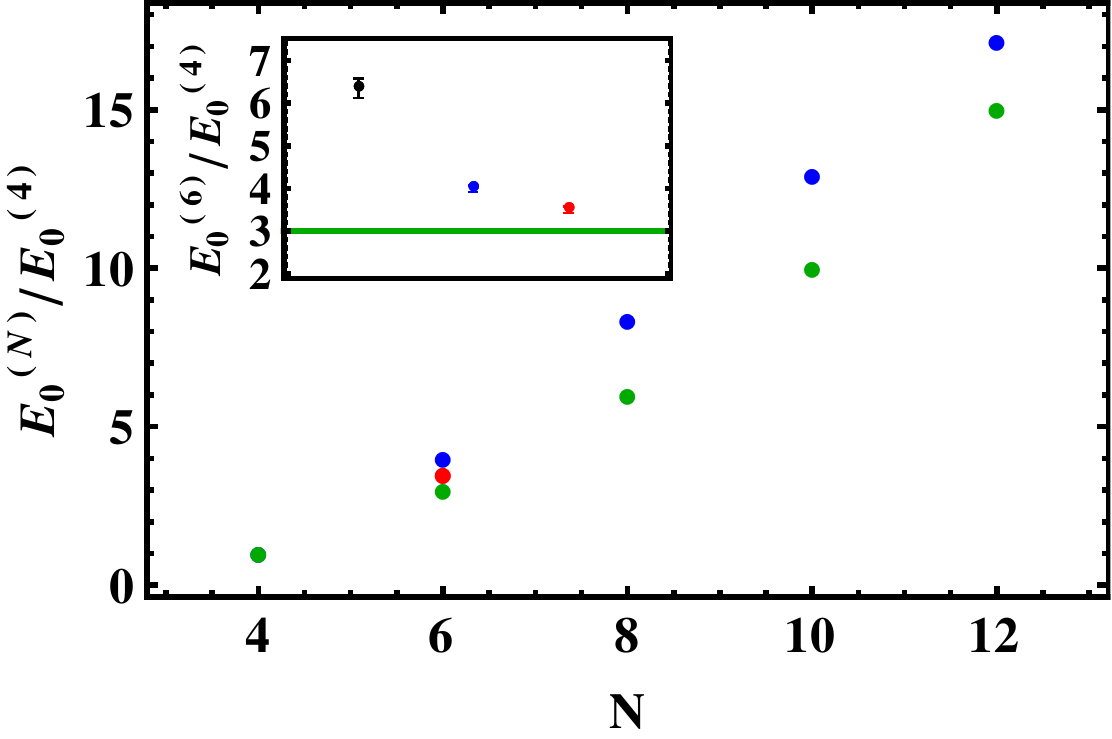}
\caption{\label{fig:comp}(Color online) Ratio of energies $E_0^{(N)}/E_0^{(4)}$ predicted by the LN distribution (green, lower points), compared to numerical calculations, (blue, upper points) \citep{vonStecher:2009qw}, and (red, $N=6$, middle point) \citep{2011PhRvL.107t0402V}. Inset: Numerical results for $E_6/E_4$ from (from left to right) \citep{2006PhRvA..74f3604H} (black), \citep{vonStecher:2009qw} (blue), and \citep{2011PhRvL.107t0402V} (red). The solid line is the result predicted by the LN distribution.}
\end{figure}

The energies for an exactly LN distribution (\Eq{LNenergies}) are plotted in \Fig{comp}, along with results from a numerical calculation of the ground state energies employing a model potential. Comparing with \citep{vonStecher:2009qw}, we find agreement at the  $10-30 \%$ level, with greater discrepancy for larger $N$. 

The inset shows the results for $E_6/E_4$ from three separate numerical calculations, \citep{2006PhRvA..74f3604H}, \citep{vonStecher:2009qw}, and \citep{2011PhRvL.107t0402V}. The three points from left to right represent a movement toward a more universal regime (see details in \citep{vonStecher:2009qw,2011PhRvL.107t0402V}). One sees a trend toward the LN result as the universal regime is reached. For $N > 6$ there is only a single ground state calculation using a finite range potential with which to compare. Because our LN predictions are based on data from a highly improved lattice theory with a very large cutoff, we predict that as studies for larger $N$ are improved and nonuniversal corrections are reduced, the resulting energies will also approach the LN results.

One may ask by how much we can deform the relation between energies and still recover a distribution that is LN in appearance. In particular, could the small discrepancy from LN seen in the lattice data give the $10-30\%$ shift in energies corresponding to those in \citep{vonStecher:2009qw}? 

To answer this question, we created mock distributions by expanding about a LN distribution and fitting the undetermined coefficients so that the moments gave the energies calculated in \citep{vonStecher:2009qw}. We created several of these distributions, using different parameterizations and fitting different combinations of energies. The results for the cumulant expansion, \Eq{cumulantexp}, of one of the mock distributions is plotted in \Fig{exp}, along with that for the lattice data of the two-body correlator. We find that the discrepancy for all mock distributions is $\sim 17\% - 30\%$, which is comparable to the difference in energies from the LN distribution and those of \citep{vonStecher:2009qw}. Furthermore, by expanding about a log-normal distribution and allowing for a $2\%$ deviation in the third cumulant, we find that the energies implied by our two-particle correlator are those of \Eq{LNenergies} to within a few percent for $N\leq 12$.

The LN distribution implies other physical consequences that connect this distribution with universal behavior. If we begin by choosing a wavefunction for our source that corresponds exactly to the solution for the two-body system, then $\mathcal{Z}_2 = 1$ and the constant of proportionality in \Eq{logmoments} is, $\mathcal{Z}_{2N} = \mathcal{Z}_4^{\frac{1}{2} N (N-1)}$. We may interpret this relation as implying that the size of the system, and therefore the overlap, scales with $N$ in the same way as the energy. 

Finally, note that if the two-body correlator is LN, then by extension the correlators for all $2N$-body Efimov states will also be LN, with rescaled parameters $\mu$ and $\sigma^2$. The logarithm of the $s$th moment for the $2N$-body correlator is given by the energy relation, \Eq{LNenergies}, $E_0^{(2Ns)} T = \frac{1}{2}N s (N s-1) E_0^{(4)} T$. Comparing to the moments of the LN distribution (\Eq{lngenmoments}), we find $\mu = \frac{1}{2} N E_0^{(4)} T \, , \sigma^2 = -N^2 E_0^{(4)} T$.

For odd numbers of particles, the correlator is not positive so a LN distribution is not expected. However, in practice it is found that these distributions are approximately LN with a small negative contribution. By fixing an additional ratio, such as $E_4/E_3$, one may extract approximate relations between the energies for odd systems. 

To summarize, a connection between the log-normal distribution and Efimov physics has been established using the distribution of the two-body correlator at unitarity. Using this connection, a novel method for obtaining the energies of the lowest of the series of $2N$-body states tied to Efimov trimers has been introduced. Lattice data strongly indicates that the distribution of this correlator is LN to within a few percent.

We note that the scaling of the energy per particle given by the LN distribution, $E_0^{(N)}/N \sim \frac{1}{4}N$, implies that cutoff independence should not hold for arbitrarily large $N$. Thus, very large moments of the distribution are not expected to conform to those of the LN distribution. However, these moments correspond to the nonuniversal regime which is inherently of less interest than the universal regime for which the LN distribution appears to be relevant. 

Given that the distribution is likely not exactly LN due to these large moments, analytical progress may be made by developing a perturbative expansion around LN, perhaps in the spirit of the semiclassical expansion introduced in \citep{Endres:2011jm}. Numerical efforts to reduce systematics in the lattice calculation of the moments of $\ln C_2(\phi)$ would help to sort out true deviations from LN from lattice artifacts, and any true deviations may be included in the overall distribution to obtain improved results for the energies. 

\acknowledgements
The author would like to thank P. Bedaque and T. Cohen for helpful discussions, and D. B. Kaplan, J. E. Drut, and J. von Stecher for comments on the manuscript. Portions of this work grew from past conversations with M. Endres, D. B. Kaplan, and J.-W. Lee. 
This work was supported by the U.S. Department of Energy under grant \#DE-FG02-93ER-40762.

\bibliography{unitary}

\end{document}